\documentclass[twocolumn,nofootinbib,showpacs,epsfig,prd]{revtex4}%
\usepackage{amsmath}
\usepackage{amsfonts}
\usepackage{amssymb}
\usepackage{graphicx}

\begin{document}

\title{Relaxation, pre-thermalization and diffusion in a noisy Quantum Ising Chain.}

\author{
Jamir Marino${}^\dag$ and Alessandro Silva${}^\ddag$}

\address{$^\dag$ SISSA – International School for Advanced Studies and INFN, via Bonomea 265, 34136 Trieste, Italy}
\address{${}^\ddag$ICTP – International Centre for Theoretical Physics, P.O. Box 586, 34014 Trieste, Italy}

\begin{abstract}

We study the dynamics of thermalization resulting from a time-dependent noise in a Quantum Ising Chain subject to a sudden quench of the transverse magnetic field. For weak noise the dynamics shows a \it pre-thermalized state \rm at intermediate time scales, eventually drifting towards an asymptotic infinite temperature steady state characterized by \it diffusive \rm behavior. By computing analytically the density of kinks, as well as the transverse and longitudinal magnetic field correlators, we characterize these two regimes, their observability and their signatures in the various physical quantities. 
\end{abstract}


\date{\today}

\maketitle

The dynamics of relaxation towards thermal equilibrium has been one of the recurrent themes of theoretical physics in the past decades~\cite{Polkovnikov2011, Lamacraft2011}. The problem is of crucial importance in many contexts, ranging from condensed matter physics to cosmology: if we  think of injecting suddenly, e.g. by an abrupt change of one of its parameters (a \it quantum quench\rm), a finite amount of energy in an otherwise closed many-body system, under which conditions will the system reach a \it thermal \rm steady state~? And how is the steady state going to be attained~? The first question has been thoroughly addressed in the literature~\cite{Polkovnikov2011, Lamacraft2011, Srednicki1994,Rigol2007}: on one hand it is natural to expect that scattering processes will in the long run lead to an ergodic, thermal redistribution of energy among the elementary degrees of freedom~\cite{Srednicki1994}. An exception are however integrable systems, where multi-particle scattering processes are highly constrained as a result of conservation laws~\cite{Rigol2007}. As recently observed in experiments with quasi-1d Bose gases,
thermalization in the usual sense will not occur~\cite{Kinoshita2006} and the asymptotic state eventually attained by the system in the thermodynamic limit is expected to be described  by an effective Generalized Gibbs ensemble (GGE) accounting for all conserved quantities~\cite{Jaynes1957,Rigol2007}.
%Far from being of just theoretical interest, the study of the dynamics of thermalization has an experimental %counterpart in the context of cold atoms in optical lattices. While early experiments on the dynamics of the Bose-%Hubbard model demonstrated beautifully many-body phase coherence through the observation of collapse and %revival phenomena, the importance of quantum integrability in the dynamics was revealed by  the lack of %thermalization in the momentum distribution of quasi-one dimensional bose gases out of equilibrium. 

While in the past few years a great deal of attention
has been paid to the description of the asymptotic steady
state, much less is known about the dynamics of equilibration
of both thermally isolated and open quantum
many-body systems. A common feature in both cases is
the expectation of a the dynamics towards the steady state is characterized
by various stages. In the case of open quantum systems,
such as spin chains coupled to classical and quantum
uniform noise ~\cite{Fubini2007,Mostame2007} or to a bosonic bath ~\cite{Patane2008,Patane2009},
these are expected to be driven by the interplay between
many-body interactions, dephasing and dissipation associated
to the external environment~\cite{Ema}. For example, the
thermalization of the system following a rapid quench of
the bath temperature is driven by the spread of thermal
correlations at a velocity set by the temperature ~\cite{Patane2009}. For
closed systems recent studies of the dynamics of quantum
field theories ~\cite{Berges2004} suggest that first the system decays to
a so-called pre-thermalized state, where the expectation
value of certain macroscopic observables is to a good approximation
”thermal”, while the distribution function
of the elementary degrees of freedom is not \cite{Berges2004,Moeckel2010}. Only
at a second stage, when energy is efficiently redistributed
by scattering processes, real thermalization occurs. Signatures
of these crossovers have been investigated theoretically
in a variety of systems (e.g. Fermi-Hubbard
models ~\cite{Moeckel2010,Eckstein2009,Kollar2011} or Spinor Condensates ~\cite{Barnet2011}), and prethermalization
has been observed experimentally in split
one dimensional condensates ~\cite{Gring2011,Kitagawa2011}. While it is evident
that the dynamics of thermalization could in general
display various crossovers, including a pre-thermalized
plateau, it is not clear from the outset whether this is a
general phenomenon for both open and closed systems, what are the conditions for its observability
on the system at hand and what are going to
be the signatures in the various observables.

In this Letter we address these issues by considering
the dynamics of thermalization of a prototypical weakly
perturbed integrable system, a Quantum Ising Chain
subject to time-dependent noise. The perturbation considered
in this work does not conserve the energy, hence
the dynamics of quantities averaged over the noise resembles
that of an open quantum system. Studying analytically
the dynamics of all essential observables and
correlation functions following a sudden quench of the
transverse magnetic field in the Quantum Ising Chain,
we will discuss the nature of the thermalization dynamics
and show that it is characterized by a crossover between
pre-thermalized and thermalized regimes. In particular,
we will show that pre-thermalization originates
from the spreading of quantum and thermal correlations
at different velocities. This effect is clearly observable for
the transverse magnetization, where it leads to a neat
crossover towards a long time diffusive behavior of the
correlators. On the other hand, it leaves much weaker
signatures on the correlation functions of the order parameter,
which has always a "thermal" form \cite{Rossini2009}. Finally
we thoroughly discuss which features of the dynamics of
this open system are expected to be relevant for the more
complex problem of thermalization resulting from integrability
breaking in a thermally isolated system.

%We will briefly discuss these results in connection to the experimental observations of Ref.[] to conclude.

Before entering technical details, let us summarize the main qualitative picture emerging from our analysis. 
We are going to study a weakly perturbed Quantum Ising chain, characterized by the hamiltononian $H=H_0+V$, where $H_0$ 
\begin{equation}
H_0=-\sum_i\sigma^x_i\sigma^x_{i+1}+g\sigma_i^z,
\end{equation}
describes the Integrable Quantum Ising chain \cite{Sachdev1999}.  Here $\sigma_i^{x,z}$ are the longitudinal and transverse spin operators at site $i$ and $g$ is the strenght of the transverse field, while 
$V=\sum_i \delta g(t) \sigma_i^z$ is a weak time-dependent white noise, with zero average and a strength
characterized by the parameter $\Gamma$, 
\begin{equation}
 \langle\delta g(t)\delta g(t')\rangle=\frac{\Gamma}{2}\delta(t-t').
\end{equation}
The Quantum Ising chain is among the simplest, yet non-trivial integrable many-body system, whose static~\cite{Sachdev1999} and dynamic properties~\cite{Rossini2009,Fagotti2011} are to great extent known. It is characterized by two dual gapped phases, quantum paramagnetic ($g>1$) and ferromagnetic ($g<1$) separated by a quantum critical point ($g=1$) where the gap $\Delta= | g-1 |$ closes. For a quench of the transverse field  all essential correlation functions have been studied extensively~\cite{Rossini2009,Fagotti2011}. In the following, we will consider the dynamics of the noisy Ising chain following a quench protocol: for times $t<0$ the system is assumed to be in the ground state of $H_0$ with $g=g_0$ and $\delta g(t)=0$, and at time $t=0$ the noise is turned on together with a global quench of the transverse field $g_0 \rightarrow g$.  We will present results for quenches within the paramagnetic phase - other types of quenches will be discussed elsewhere~\cite{Marino2012}.

%Local operators in the quasiparticles, such as transverse magnetization, are qualitatively sensitive to the details% %of the integrability of the model, so in our case the correlator of the transverse magnetization develops a %%\emph{diffusive} behaviour due to the noise perturbation. Non local operators, like the order parameter, show an %exponential decay common to a variety of quenches (see for instance \cite{Rossini} and \cite{Levitov}), although %the exponent is dictated by the relevant scales of the model, $\Gamma t$ and $g$.

\begin{figure}[htbp]
\centering
\includegraphics[scale=0.8]{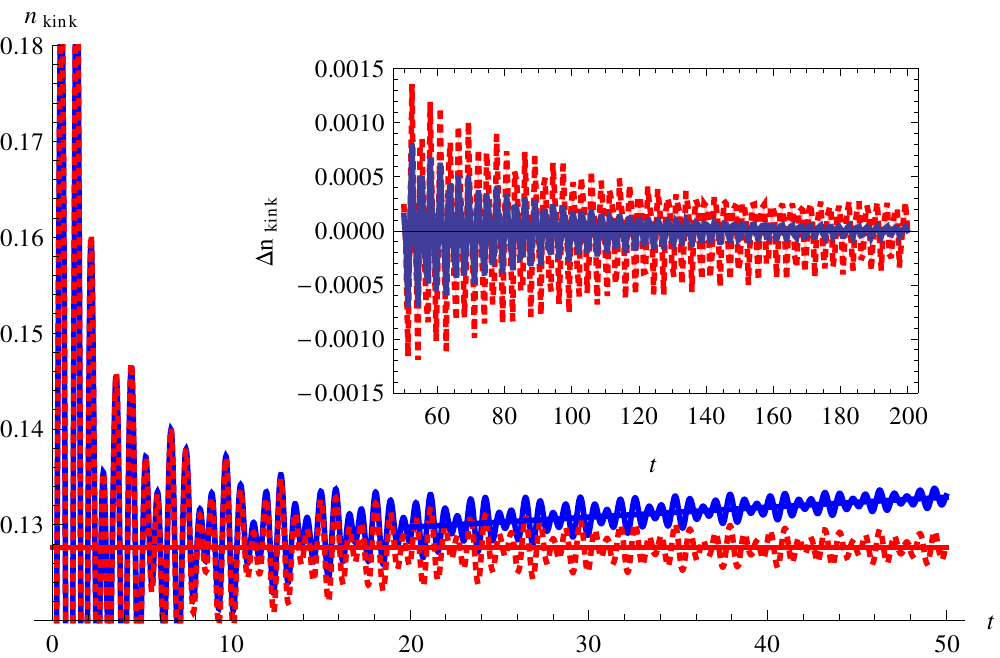}
\caption{[\it Color online \rm]  The density of kinks vs. time for a quench with $\Gamma=0.01$, $\Delta=3$, $\Delta_0=0.1$ (blue full line). While the red dotted curve is the value attained by $n_{\rm kink}$ without perturbation (the red horizontal straight line sets the asympthotically value predicted by GGE),  the full blue line shows the time evolution with noise that first appears to saturate at the GGE value, but later runs away towards the infinite temperature state (the blue oblique straight line is $n_{drift}$). [\it Inset \rm] $\Delta n_{kink}$ vs. time. The dephasing of the quantum oscillations is accelerated in the presence of the noise (blue full line) in comparison with the case without perturbation (red dotted line).}\label{kinks}
\end{figure}

A qualitative picture of the mechanism of pre-thermalization can be obtained by looking at the separation of time scales associated to three distinct physical effects \cite{Kollath}.  First of all, the \it coherent \rm superposition of modes with different frequencies leads to a first, power law decay of physical quantities towards the pre-thermalized (yet non-thermal) state described by a GGE~\cite{Berges2004,Kollar2011,noteGGE}. This effect is similar to inhomogeneous broadening. This first decay is abruptly accelerated on time scales of order of $1/\Gamma$ by the intervention of \emph{noise induced dephasing}. Finally, on much longer times scales the occupation of quasi-particles evolves towards its thermal value (in our case corresponding to infinite temperature).  All of this is clearly observed in the time evolution of the density of kinks $n_{\rm kink}(t)=\langle \sum_i (1-\sigma_i^x \sigma^x_{i+1})/2  \rangle$, which can be written as $n_{kink}(t)=n_{drift}(t)+\Delta n(t)$; here, $n_{drift}$ describes the heating of the system towards the state of infinite temperature, and for finite $\Gamma$ and long times is given by $n_{\rm kink}\simeq\frac{1}{2}-\frac{m_1}{4\sqrt{\pi}\sqrt{\Gamma t}}$, while $\Delta n(t)$ describes the sharp oscillations related to the quantum evolution induced by the quench (see Fig.~\ref{kinks} and discussion after Eq.(16)). 

This sequence of crossovers can be further analyzed  by studying the spreading of correlations  $\rho_{\alpha}(r,t)=\langle \sigma^{\alpha}_i(t) \sigma^{\alpha}_{i+r}(t) \rangle$ in the transverse ($\alpha=z$) and longitudinal ($\alpha=x$) directions. A pre-thermalized plateau emerges when  quantum and thermal correlations propagate at sufficiently different velocities: this is manifest in the various regimes of the transverse correlation functions. The first crossover is observed for $\Gamma t\ll1$
\begin{equation}
\rho^{zz}(r,t)\simeq_{\Gamma t\ll1}\begin{cases}
\frac{1}{2\pi r^2}\exp[-2\Delta_0r] & \text{$r\gg t$}\\
& \\
\frac{1}{2\pi r^2}\exp[-r/\xi_z] & \text{$r\ll t$}
\end{cases}
\end{equation}
where $\xi_z$ is the correlation lenght associated to the \emph{quantum quench protocol}.
For larger times, namely $\Gamma t\gg1$, the noise becomes relevant and the second crossover, between exponential and \it diffusive \rm behavior of the correlator, emerges
\begin{equation}
\label{corto}
\rho^{zz}(r,t)\simeq_{\Gamma t\gg1}\begin{cases}
\frac{1}{2\pi r^2}\exp[-r/\xi_z] & \text{$\gamma t\ll r\ll t$}\\
& \\
-\frac{1}{\pi}\frac{\Delta^2}{4}\frac{1}{\Gamma t}\exp\left[-\frac{(\Delta r)^2}{2\Gamma t}\right] & \text{ $r\ll \gamma t$}
\end{cases}
\end{equation}
with $\gamma=\Gamma/\Delta$. The correlation length, which is dictated at $t \ll r$ by the initial gap $\Delta_0$, crosses over as  $r \simeq t$ to an intermediate asymptotic form depending by the masses of the quench, through the function $\xi_z=\xi_z(g_0,g_1)$ ~\cite{Fagotti2011}. Though admittedly \it thermal \rm behavior is stronger in the longitudinal correlators (see below), the state reached in the intermediate plateau is the one the system would have reached in the absence of noise, hence in a sense the pre-thermalized state of the weakly perturbed integrable model~\cite{Kollar2011,Kitagawa2011}.
The first crossover is driven by the light-cone effect ~\cite{Polkovnikov2010,Calabrese2006}: this intermediate plateau persists until a second front, propagating at speed lower by a factor $\gamma$ and carrying thermal correlations, passes through $r$. At this point the correlator crosses over to a \it diffusive \rm time-dependent form, consistent with thermal Glauber dynamics \cite{Glauber1963}, indicating the continuous heating of the system towards the infinite temperature state. A similar behaviour emerges also for quenches across the critical point or starting at the critical point.
In the absence of the quench, the correlator doesn't show the intermediate prethermalized regime and the equilibrium correlator -$e^{-2\Delta_0r}$- crosses over directly to the diffusive behaviour \cite{Marino2012}.
\begin{center}
\begin{figure}[htbp]
\includegraphics[scale=0.37]{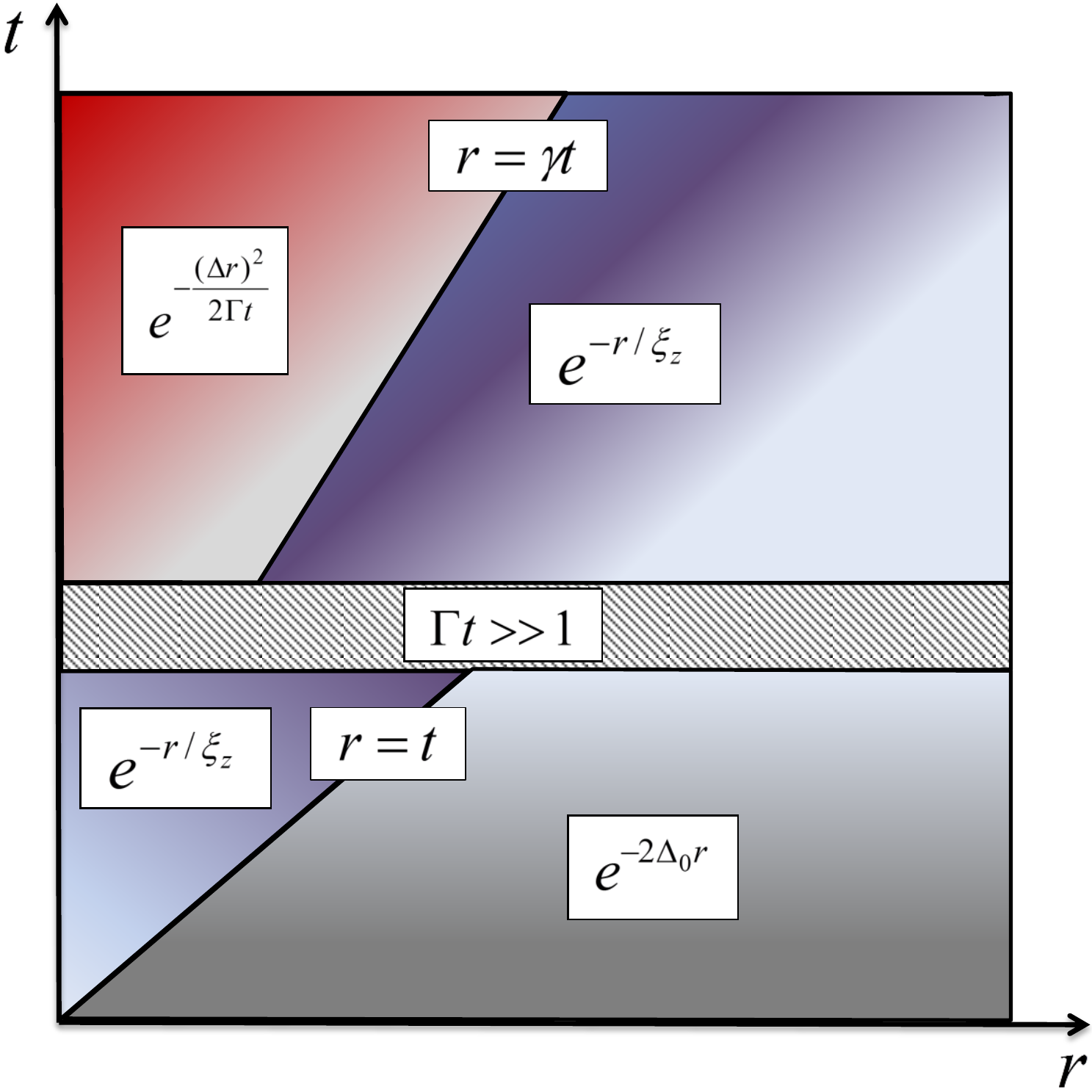}
\caption{The spreading of quantum and thermal correlations in the noisy Quantum Ising Model: the transverse field correlator has a first crossover when ballistic quasi-particles, carrying quantum correlations, propagate at the distance $r$. Thermal correlations propagate at a second stage, leading to a crossover to a diffusive form, consistent with thermal Glauber dynamics. }
\end{figure}
\end{center}
\emph{Thermal} behavior in the intermediate plateau is observed more clearly in the correlation function of the longitudinal magnetization. For a quench without dissipation this correlator  has in general an exponential form
%\begin{eqnarray}
$\rho^{xx}(r,t)\sim \exp[- r/\xi]$,
%\end{eqnarray} 
with a correlation length $\xi$ dictated by the non-thermal distribution function of quasi-particles and predicted by the Generalized Gibbs ensemble \cite{Fagotti2}. For small quenches, however, it turns out that $\xi$ can be efficiently parameterized in terms of an effective temperature $T_{\rm eff}$~\cite{Rossini2009} in the form $\xi\simeq\sqrt{\pi/(2T_{\rm eff} \Delta)}\;\exp[\Delta/T_{\rm eff}]$. The emergence of this \it thermal behavior \rm (not to be confused with thermalization) in a system with a non-thermal distribution of quasi-particles is strikingly similar to the phenomenology of pre-thermalization observed experimentally in split condensates~\cite{Gring2011} (the two effects turn out to have a common physical origin~\cite{Menegoz2011}). Turning on the noise, the signatures of the crossover observed for the transverse magnetization are expected in this case to be different. Indeed,  computing the correlation function following a switching on of the noise starting with the equilibrium state with gap $\Delta_0$ one obtains
\begin{equation}\label{fuori}
\rho^{xx}(r,t)\sim r^{-1/2}\;e^{-r/\xi(t)}
\end{equation}
where $1/\xi(t)=(1/\xi+\frac{\Gamma t}{2(1+\Delta_0)^2})$ and $1/\xi=\log(1+\Delta_0)$. Since the same exponential form persists  the spreading of quantum and thermal correlations will not result in a diffusive form, but rather modify just the specifics of the correlation length which at later times shrinks as $1/\Gamma t$. 
The different signatures observed in the transverse and longitudinal magnetization are consistent with analogous effects observed elsewhere~\cite{Rossini2009,Canovi2011}.

Let us now present some of the technical details of our analysis. Our main task will be to write a closed kinetic equation for the quasi-particles of the Quantum Ising chain subject to the influence of noise. %First we explain how Keldysh formalism -developed for the treatment of out of equilibrium interacting many body %systems \cite{Keldysh} (for a review see \cite{Landau} or \cite{Kamenev})-  allows us to write a master equation %for our system under the effect of the noise. Then we are going to elucidate the quench protocol and we will %discuss the results for the transverse magnetization correlator, showing its causality diagram and under which %conditions \emph{diffusive} behaviour arises. Finally, we show how the relaxation exponent of the order %parameter correlator can be extracted from theorems on the asymptotics of Toeplitz determinants. 
Quasi-particles can be introduced in terms of Jordan-Wigner fermions \cite{Sachdev1999}, $c_i$, through the relation $\sigma_i^z=1-2c_i^\dag c_i$ and $\sigma_i^x=-\prod_{j<i}(1-2c_j^\dag c_j)(c_i+c_i^\dag)$. The Hamiltonian takes in Fourier space, $c_k=\sum_jc_je^{ikj}$, the simple form
\begin{equation}
H=2\sum_{k>0}\widehat{\psi}_k^\dag \widehat{H_k}\widehat{\psi}_k
\end{equation}
where
\begin{equation}
 \widehat{H_k}=\widehat{H_k^0}+\delta g(t)\sigma_z
\end{equation}
$\widehat{H_k^0}=(g-\cos{k})\sigma_z-(\sin{k})\sigma_y$, $\widehat{\psi}_k$ is the Nambu spinor $\bigl(\begin{smallmatrix}c_k\\ c^\dag_{-k}\end{smallmatrix} \bigr)$, and $\sigma_y$, $\sigma_z$ are the Pauli matrices in the 2$\times$2 Nambu space. Without noise, the diagonal form $H=\sum_{k>0}E_k(\gamma^{\dag}_k\gamma_k-\gamma_{-k}\gamma^{\dag}_{-k})$, with energies $E_k=\sqrt{(g-\cos k)^2+\sin^2k}$, is achieved after a Bogoliubov rotation $c_k=u_k(g)\gamma_k-iv_k(g)\gamma_{-k}^{\dag}$ and $c_{-k}^{\dag}=u_k(g)\gamma_{-k}^{\dag}-iv_k(g)\gamma_{k}$; the coefficients are given by 
\begin{equation}
u_k(g)=\cos(\theta_k(g)) \qquad v_k(g)=\sin(\theta_k(g))
\end{equation}
where $\tan(2\theta_k(g))=\sin(k)/(g-\cos(k))$. Notice that the gap in the spectrum is  $\Delta=|g-1|$ \cite{Sachdev1999}.

%\begin{figure}[htbp]
%\centering
%\vspace{0.2cm}
%\includegraphics[scale=0.3]{Born.eps}
%\caption{Self-energy diagrams in the Born approximation. The dash-dotted lines represent averages over the %noise, the full line the bare quasi-particle propagators, and the double line the full double line the dressed quasi-%particle propagator. }
%\end{figure}

%At $t=0$ our system is prepared in the ground state of the Ising model $\arrowvert\psi_0(g_0)\rangle$ and we %let it evolve under the quantum Ising hamiltonian with transverse field $g$, coupled to a delta correlated classical %noise of amplitude $\Gamma$
%\begin{equation}\label{Hnoise}\begin{split}
% H=\sum_i\sigma_i^x\sigma_{i+1}^x&+g\sigma_i^z+\delta g(t)\sigma_i^z\\
%\end{split}\end{equation}

In order to describe the dynamics of the Ising model under the effect of the noise we will now derive a set of 
kinetic equations for the lesser Green function \cite{Kamenev2012}
\begin{equation}
 G^<(t,t')=\Big[G_k^<(t,t')\Big]_{i,j}=i\langle\psi_{k,j}^\dag(t')\psi_{k,i}(t)\rangle,
\end{equation}
in terms of which we will express all physical observables of the model.
%We start recalling \cite{Kamenev} the definition of the statistical Green function on the Keldysh contour 
%\begin{equation}
% G^c=-i\langle T_c\psi_{ki}(t)\psi^\dag_{kj}(t')\rangle
%\end{equation}
%where $T_c$ is the time ordering operator on the Keldysh contour \cite{Kamenev}, and defining the 
%Following \cite{Patane} 
Our kinetic equation is encoded in the Dyson equation for the contour ordered Green's function
\begin{equation}
 G^c_{t,t'}=G_{0_{t,t'}}^c+G_{0_{ t,t''}}^c\Sigma^c_{t'',t'''}G^c_{t''',t'},
\end{equation}
where $G_{0_{t,t'}}^c$ is the unperturbed Green function and $\Sigma^c_{t,t'}$ is the self energy;  in right hand side it is understood a convolution product, all the quantities are evaluated along the Keldysh contour. In the following we will take all self-energies within the \emph{self-consistent Born approximation} \cite{Patane2008}, controlled by the small parameter $\gamma$.
%The Dyson equations for the statistical and advanced Green function is
%\begin{equation}\label{dyson}
%\begin{split}
% i\partial_tG^<(t,t')=&H_kG^<(t,t')+\int dt'' [\Sigma^<(t,t'')G^a(t'',t')+\\
%& +\Sigma^r(t,t'')G^<(t'',t')]\\
% -i\partial_{t'}G^<(t,t')=&G^<(t,t')H_k+\int dt'' [G^r(t,t'')\Sigma^<(t'',t')+\\
%& +G^<(t,t'')\Sigma^a(t'',t')]
%\end{split}
%\end{equation}
Specializing now the Dyson equation to the lesser and advanced  Green's function and defining the equal time matrix $\rho_t\equiv-iG^<_{t,t}$, we obtain with simple algebraic manipulations 
\begin{equation}\label{master2}
\partial_t\rho_k=-i[H_k^0,\rho_k]+\frac{\Gamma}{2}(\sigma\rho_k \sigma-\rho_k),
\end{equation}
the second term in the right hand side takes into account the effect of the noise, and $\sigma\equiv\cos2\theta_k\sigma_z+\sin2\theta_k\sigma_y$ \cite{noteC}. Here $\rho_k$ are expressed in the basis of the fermions diagonalizing $H_0$
\begin{equation}
\rho_k= \begin{pmatrix}
 \langle\gamma^{\dag}_{k}\gamma_{k}\rangle & \langle\gamma^{\dag}_{k}\gamma^{\dag}_{-k}\rangle \\
 \langle\gamma_{-k}\gamma_{k}\rangle & \langle\gamma_{-k}\gamma^{\dag}_{-k}\rangle
 \end{pmatrix}
\end{equation}
where $\langle\gamma^{\dag}_{k}\gamma_{k}\rangle$ are the \emph{populations} of levels of momentum $k$ and $\langle\gamma^{\dag}_{k}\gamma^{\dag}_{-k}\rangle$ the \emph{coherences}. While this kinetic equation is analogous to a Lindblad master equation for a two level system, the Keldysh diagrammatic technique has the advantage of clarifying the approximations made in a language appropriate, and easily generalized to other many-body problems~\cite{Kamenev2012}. 
% the system is in the ground state of the paramagnetic phase ($g_0>1$), $|\psi(g_0)\rangle\equiv|\psi_0\rangle$ %and we let the system evolve under the combined action of the new Hamiltonian and of the noise. 
%\begin{figure}[htbp]
%\centering
%\includegraphics[scale=0.9]{Relaxation.eps}
%\caption{[\it Colors online \rm] The occupation of quasi-particles $n_k=1/2+\delta f_k$ vs. $k$ obtained by %solving Eq.(\ref{sistema1})-(\ref{sistema3}) starting with the ground state at $g=1.5$ (here $\Gamma=1$) for %various times. Notice that the modes close to $k=0,\pi$ have a much slower relaxation than those at the centre %of the band.}\label{Relaxation}
%\end{figure}%The knowledge of  the matrix $\rho_k$ allows the direct computation of the density of the kinks as well as equal %time correlation functions of the transverse and longitudinal magnetization \cite{Sachdev1999}, but also the %number of quasiparticles produced after the quench, as we will point out in \cite{Marino2012}. 
Parameterizing the matrix $\rho_k$ in the form
%\begin{equation}\label{ansatz}
$\rho_k=\frac{1}{2}\textbf{1}+\delta f_k\sigma_z+x_k\sigma_x+y_k\sigma_y$,
%\end{equation}
we obtain a set of three coupled equations. 
\begin{eqnarray}\label{sistema1}
\partial_t(\delta f_k)&=&-\Gamma\sin^22\theta_k\delta f_k+\frac{\Gamma}{2}y_k\sin4\theta_k,\\
\partial_t x_k&=&-\Gamma x_k-2E_ky_k,\label{sistema2}\\
\partial_t y_k&=&\frac{\Gamma}{2}\sin4\theta_k\delta f_k+2E_kx_k-\Gamma\cos^22\theta_k y_k. \label{sistema3}
\end{eqnarray}
The initial time condition depends of course on the physical situation to be studied: for a quench from $g_0>1$ to $g>1$, one obtains $\rho_k=1/2+(\cos(2\Delta \alpha_k)/2)\;\sigma_z +
(\sin(2\Delta \alpha_k)/2)\; \sigma_y$ (where $\Delta\alpha_k=\theta_k(g)-\theta_k(g_0)$).
These equations can be readily solved exactly and the resulting long time dynamics turns out to be dominated by the modes close to $k=0, \pm \pi$ 
This can be clearly evinced by neglecting the terms coupling $\delta f_k -y_k$ in Eq.(\ref{sistema1})-(\ref{sistema3}), in the limit $\gamma\ll1$. In that case Eq. \eqref{sistema1} gives
%\begin{equation}\label{esatta}
$\delta f(t)=(\sin^2(\Delta\alpha_k)-1/2) e^{-\Gamma\sin^22\theta_1t}$,
%\end{equation}
where $\Delta\alpha_k=\theta_1-\theta_0$. 
%computing the dynamics of the occupation of quasi-particles $n_k=1/2+\delta f_k$ (see Fig.(\ref{Relaxation})). %Starting for example from the zero temperature case, 
In this expression we can now see that the relaxation rates tend to vanish close to the band edges ($k=0\pm \pi$),
while most of the modes relax fast to their thermal occupation ($n_k \simeq 1/2$) on time scales of the order of $1/\Gamma$. %In order to make the analysis physically transparent, it is convenient to express the base $g$ in terms of the %ones in the base $g_0$, according to \cite{Silva} \cite{SGambassi} and, using master equation \eqref{master2} 
The coherences in turn decay exponentially fast: focusing on $k \simeq 0$, and using $\gamma\ll1$, we indeed find 
\begin{equation}\label{decadimento}\begin{split}
\delta f_k&=\frac{1}{2}\Big(\frac{k^2}{2m^2}\rho_{-}^2-1\Big)e^{\frac{-\Gamma k^2t}{m^2}}\\
x_k-iy_k(t)&=-\frac{ik}{2m}\rho_{-}e^{-\alpha t-i\beta t}\\
\end{split}
\end{equation}
where we have defined $\rho_-\equiv (\Delta_0-\Delta)/\Delta_0$ \cite{nota1}, $\alpha\simeq \Gamma +O(k^2)$ and 
$\beta\simeq 2 \Delta + O(k^2)$. A similar analysis expansion is of course possible close to $k=\pm \pi$.

%$\alpha\equiv-\Gamma\Big(1-\frac{1}{2}\Big(\frac{k}{m}\Big)^2\Big)$ and
%$\beta\equiv2 m\Big(1+\frac{1}{2}\Big(\frac{k}{m}\Big)^2\Big)$.

%\bf time evolution of occupations and coherences ... Density of quasi-particles and energy. \rm

%In this section we compute the number of defects produced after the quench, defined as
%\begin{equation}
%N\equiv\frac{1}{2}\sum_i(1-\sigma_i^x\sigma_{i+1}^x)
%\end{equation}

Using the above expressions for the populations and the coherences, the computation of the number of kinks and of the correlators follow the standard tecniques used for the Quantum Ising Model \cite{Rossini2009, Calabrese2006, Sachdev1999, Fagotti2, McCoy, Ovchi, FH} and the results shown in the first part of the paper comes after ordinary algebraic manipulations, that we will report elsewhere \cite{Marino2012}. For instance, in the number of kinks, $n_{kink}=n_{drift}(t)+\Delta n(t)$, where $n_{drift}(t)=\sum_{k>0}\frac{1}{2}+\delta f_k\cos2\Delta\alpha_k^*$ and $\Delta n(t)=\sum_{k>0} \sin2\Delta\alpha_k^*y_k$, the first term describes the drift towards the infinite temperature state and the second one, due to the coherences, describes oscillations suppressed by the noise on a time scale $\sim 1/\Gamma$ (see inset of Fig. 1).

In conclusion, we have discussed the dynamics of thermalization in a weakly perturbed integrable model, a noisy Quantum Ising chain. Though the results presented in this work pertain to a noisy, open quantum system, we believe that some of the features discussed, such as crossovers between prethermalized and thermal regimes driven by the spreading of fronts at different velocities,
will be observed also in the dynamics of thermalization of weakly non-integrable models. Similarly, we do expect different signatures of thermalization to be observed consistently in the transverse and longitudinal correlators \cite{Rossini2009,Canovi2011}. At the
present, however, it is not clear whether a weak breaking of integrability in the Ising model will generically lead to diffusive behaviour
of the transverse magnetization correlator: this issue remains to be addressed by future studies.

%of using Kelysh formalism it has been possible to write a master equation which yelds the time evolution in term %of a Dyson equation for an Ising chain whose integrability has been broken by a classical random field. Written in %the Bogoliubov basis, the master equation allows us to compute the correlators of relevant observables of the %model for long times, $\Gamma t\gg1$; our results show for the $m^z$ correlator a slow diffusion wave front, %associated to the classical noise, and a fast wave front associated to the effects of the quantum quench. On the %other side, order parameter correlator doesn't develop a diffusive behaviour and we interpret this result as a %signature of the non locality of the order parameter in the quasiparticles diagonalizing the starting Hamiltonian.

%\section{Number of defects}
We would like to thank G. Biroli, P. Calabrese, M. Fabrizio, R. Fazio, A. Gambassi, G. Mussardo, C. Kollath, G. Santoro,  and in particular M. Marcuzzi for useful and enlightening discussions. A. S. would like to thank the Galileo Galilei Institute in Firenze for hospitality during the completion of this work.


\begin{thebibliography}{30}
\bibitem{noteC}
The noise can introduce correlations between modes of different $k$ but they can be neglected in the limit $\gamma\ll1$.
\end{thebibliography}
\end{document}